 \documentclass[smallabstract,smallcaptions]{dccpaper}

\usepackage{epsfig}
\usepackage{citesort}
\usepackage{amsmath}
\usepackage{amssymb}
\usepackage{color}
\usepackage{url}
\usepackage{subfigure}
\usepackage{booktabs}
\usepackage{caption}
\usepackage{comment}

\newlength{\figurewidth}
\newlength{\smallfigurewidth}

\begin{document}

\title
{\large
\textbf{Learning Optimal Linear Block Transform by Rate Distortion Minimization} 
}

\author{%
Alessandro Gnutti$^{\ast}$, Chia-Hao Kao$^{\ast}$, Wen-Hsiao Peng$^{\dag}$, and Riccardo Leonardi$^{\ast}$\\[0.5em]
{\small\begin{minipage}{\linewidth}\begin{center}
\begin{tabular}{ccc}
$^{\ast}$University of Brescia & \hspace*{0.1in} & $^{\dag}$National Yang Ming Chiao Tung University \\
Department of Information Engineering && Department of Computer Science \\
Brescia, Italy && Hsinchu, Taiwan\\
\end{tabular}
\end{center}\end{minipage}}
}

\maketitle
\thispagestyle{empty}

\begin{abstract}
Linear block transform coding remains a fundamental component of image and video compression. Although the Discrete Cosine Transform (DCT) is widely employed in all current compression standards, its sub-optimality has sparked ongoing research into discovering more efficient alternative transforms even for fields where it represents a consolidated tool.
In this paper, we introduce a novel linear block transform called the Rate Distortion Learned Transform (RDLT), a data-driven transform specifically designed to minimize the rate-distortion (RD) cost when approximating residual blocks. Our approach builds on the latest end-to-end learned compression frameworks, adopting back-propagation and stochastic gradient descent for optimization. However, unlike the nonlinear transforms used in variational autoencoder (VAE)-based methods, the goal is to create a simpler yet optimal linear block transform, ensuring practical integration into existing image and video compression standards.
Differently from existing data-driven methods that design transforms based on sample covariance matrices, such as the Karhunen-Loève Transform (KLT), the proposed RDLT is directly optimized from an RD perspective. Experimental results show that this transform significantly outperforms the DCT or other existing data-driven transforms. Additionally, it is shown that when simulating the integration of our RDLT into a VVC-like image compression framework, the proposed transform brings substantial improvements. All the code used in our experiments has been made publicly available at \cite{github}.
\end{abstract}

\Section{Introduction}
\label{sec:intro}
Image and video compression focuses on encoding data efficiently while preserving visual quality. Traditionally, linear block transform coding has been a cornerstone of such compression techniques. The key principle behind this approach is to use transforms that produce statistically independent coefficients in the transform domain, allowing more efficient encoding.

The 2D Type-2 Discrete Cosine Transform (DCT-II, or simply DCT) has emerged as the most widely used transform over the past few decades, and it has been integral to many block-based compression techniques. It has been employed in video compression standards from ITU/H.261 \cite{Bhaskaran1995} through JVT/H.265 (HEVC) \cite{intraHEVC2012} and the latest JVT/H.266 (VVC) \cite{vvc2021}. Similarly, in image compression, DCT is foundational, from the long-standing ISO/JPEG standard \cite{jpeg1992} to the more recent BPG format \cite{bpg}, which is based on HEVC intra-frame coding.

The popularity of the DCT originally stems from its ability to approximate the Karhunen-Loève Transform (KLT) under first-order Markov assumptions \cite{Rao1975}. While these assumptions align well with the statistical properties of natural images, modern coding systems apply transforms to residuals from prediction processes, which have different statistical characteristics than the pixel domain, breaking its direct relationship with the KLT.

The search for alternative, more efficient transforms has long been a compelling area of research. Most previous studies have focused on designing KLTs based on sample covariance matrices derived from specific training sets. However, the quantization process used for entropy coding the transform coefficients more closely resembles a non-linear approximation of the image block, since the $n$ most significant quantized coefficients are selected to represent the image block. In contrast, the KLT is optimal for linear approximation, in which the first $n$ coefficients with the largest variance are used. Consequently, the theoretical optimality of the KLT does not fully hold in practical compression systems.

The recent surge in deep learning has catalyzed a new era of advancements in image compression, with end-to-end learned systems garnering significant attention. Notably, variational autoencoder (VAE)-based methods \cite{kingma2022autoencodingvariationalbayes} have achieved image compression performance that closely rivals the latest VVC intra coding. Unlike traditional hand-crafted codecs, VAE-based approaches typically employ a nonlinear image-level transform, which converts an input image into a compact set of latent features, significantly reducing the dimensionality compared to the original image.

However, the implementation of end-to-end learned compression systems is still far from being adopted in standard compression frameworks. This is primarily due to the need for a major overhaul of existing hardware that currently supports traditional compression methods. Additionally, the computational complexity—particularly on the decoder side—remains significantly higher compared to conventional codecs. Therefore, improving the performance of traditional codecs by searching for more efficient linear block transforms remains a highly active area of research.

In this paper, we introduce a novel linear block transform, referred to as the Rate Distortion Learned Transform (RDLT). RDLT is a data-driven transform specifically designed to minimize the rate-distortion (RD) cost when approximating residual blocks.
Drawing from the foundational work \cite{ballé2017endtoendoptimizedimagecompression}, which laid the groundwork for all the end-to-end learned compression frameworks, our method leverages learning frameworks similar to those employed in end-to-end learned compression systems. However, unlike nonlinear transforms in VAE-based methods, the goal is to design a simpler, yet optimal, linear block transform, allowing it to be integrated into current image/video compression standards.

To the best of our knowledge, this is the first attempt to learn an optimal linear block transform using a procedure commonly employed in end-to-end learned methods. Specifically, the proposed approach enables accurate estimation of the rate for representing the coefficients of the designed transform adopting a parametric Gaussian entropy model.
Experimental results demonstrate superior performance to the DCT and other competitive data-driven transforms. When simulating the integration of RDLT into a VVC-like image compression framework, the proposed transform leads to significant performance improvements.

The main contributions of this work are:
\begin{itemize}
    \item A novel linear block transform specifically designed to minimize the RD cost when approximating residual blocks.
    \item Although the method leverages the advanced principles commonly used in end-to-end learned approaches, the resulting transform allows practical integration into existing image and video compression standards.
    \item Experimental results demonstrate that the proposed transform significantly outperforms both DCT or other existing data-driven transforms. Also, when simulating the integration of RDLT into a VVC-like image compression framework, the results show substantial improvements over the current VVC setting.
\end{itemize}

The reminder of the paper is organized as follows: related work is first reviewed. Then, the proposed method is extensively described, outlining the process for generating the RDLT; experimental simulations validate the resulting performance. Finally, conclusions are drawn.

\Section{Related work}
\label{sec:related}


In recent image and video compression systems, the sub-optimality of the DCT has been overcome by integrating a Rate-Distortion Optimized Transform (RDOT) scheme \cite{zhao2011video}. In this context, the encoder and decoder share multiple transforms. The encoder then selects the optimal transform for each block by minimizing the corresponding RD cost. In particular, the authors of \cite{zhao2011video} propose incorporating a set of offline, data-driven KLT transforms into the compression framework.

When applying the transform to residual blocks, another possible approach to improving transform efficiency is to tailor transforms to the specific statistics of residual data. For example, Mode-Dependent Transform (MDT) schemes, which design a unique KLT for each intra-prediction mode, have been proposed for both H.264/AVC \cite{ye2008improved} and H.265/HEVC \cite{takamura2013intra}\cite{arrufat2014non}. 

However, as previously noted, the optimality of the KLT is only guaranteed when used for linear signal approximation. Since compression is a form of non-linear approximation, the KLT's optimality cannot be assured in this context. 
In light of this, the latest compression standards \cite{intraHEVC2012} \cite{vvc2021} do not integrate KLT-based solutions, but adopt a RDOT scheme including simple DCT variants. In HEVC, the Discrete Sine Transform (DST) supplements the DCT, while its successor VVC includes four additional primary transforms besides the DCT-II: the type-$8$ DCT (DCT-VIII), the type-$7$ DST (DST-VII), and their horizontal and vertical combinations.

Recent research has also explored the use of Graph Fourier Transforms (GFTs) for image and video coding, building on the principles of Graph Signal Processing (GSP) \cite{Ortega-proceedings-2018}. In GSP, graphs are used as models where vertices carry signal samples, and edge weights typically convey existing similarity between them. In \cite{fracastoro2019graph}, a block-adaptive scheme for image compression with GFTs is proposed, where a graph is constructed for each block by minimizing a regularized Laplacian quadratic term as a proxy for the actual RD cost; this approach remains suboptimal. Other works propose GFTs tailored to specific image characteristics \cite{wei2015} \cite{egilmez2020} or leverage the symmetric properties of graphs \cite{gnutti18} \cite{Gnutti_icip19} \cite{gnutti21}.

The approach in \cite{sezer2015sparse} attempts to overcome the KLT's limitations in nonlinear approximation by introducing a data-driven framework that employs Sparse Orthonormal Transforms (SOTs). These transforms are optimized by minimizing a cost function consisting of a distortion term and a rate term, which is approximated as the $l_0$ norm of the transform coefficients. 
However, we will show that accurate rate estimation is crucial, and that the proposed RDLT, designed with a learning-based image compression system associated to a parametric Gaussian entropy model for rate approximation, surpasses all existing linear block transforms in terms of performance.


\Section{Proposed method}
\label{sec:method}

The proposed Rate Distortion Learned Transform (RDLT) is designed as a linear block transform learned through back-propagation in an end-to-end manner. By designing a linear transform that adapts to the statistical characteristics of residual blocks from an RD perspective, the aim is to achieve higher coding efficiency compared to traditional transforms such the DCT, or other data-driven transforms.

By adopting a framework similar to the one proposed in ~\cite{ballé2017endtoendoptimizedimagecompression,ballé2018variational}, the approach involves training a single linear block transform using a rate estimation model which targets a given RD objective. It wants to maintain compatibility with existing block transform-based standards, thus facilitating its future integration.

\begin{figure}
    \centering
    \includegraphics[width=1\linewidth]{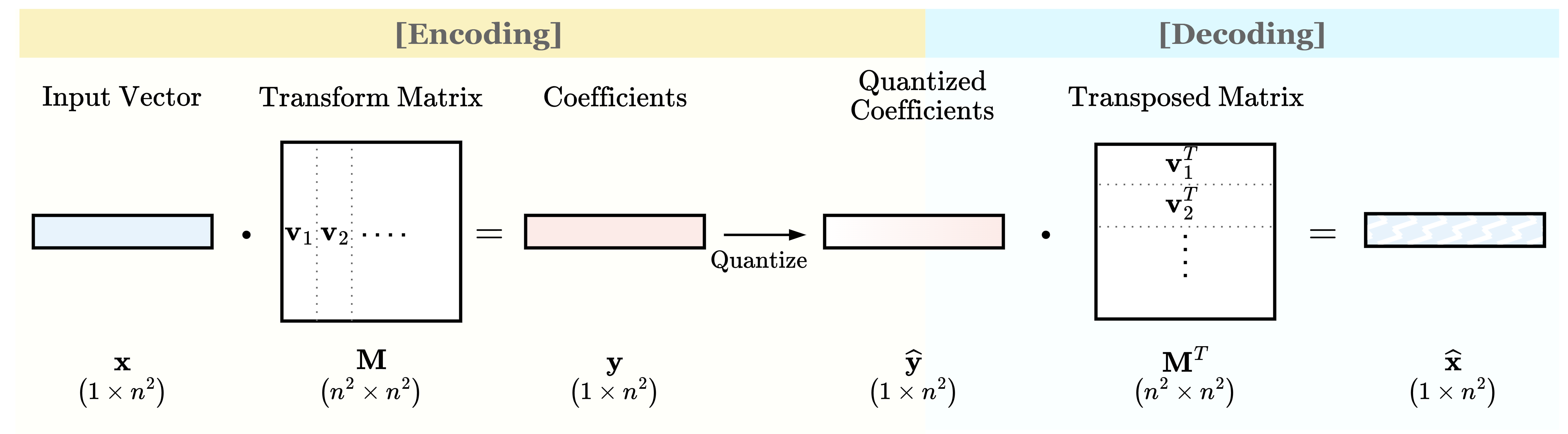}
    \caption{Overall pipeline for linear block transform coding. The image block is represented as a row vector.}
    \label{fig:framwork}
\end{figure}

\SubSection{Problem formulation}
Figure~\ref{fig:framwork} illustrates the typical steps in a linear block transform coding framework, including transforming and quantizing an input residual block of size $n \times n$, along with the inverse operations for reconstruction at the decoding stage. The input block is reshaped to form a row vector $\mathbf{x}\in \mathbb{R}^{1\times n^2}$, and the transform matrix $\mathbf{M}$ is viewed as a set of $n^2$ column vectors $\mathbf{v}_1,\mathbf{v}_2,\ldots,\mathbf{v}_{n^2}$ serving as basis functions. For a given $\mathbf{x}$, the transform is applied as:
\begin{equation}
    \mathbf{y} = \mathbf{x}\mathbf{M},
\end{equation}
where $\mathbf{y}\in \mathbb{R}^{1\times n^2}$ denotes the transformed coefficients. These coefficients are subsequently quantized to obtain the quantized coefficients $\hat{\mathbf{y}}$, using scalar quantization for a given step-size $Q$:
\begin{equation}
    \hat{\mathbf{y}} = \text{round}(\frac{\mathbf{y}}{Q}).
\end{equation}
The quantized coefficients are then entropy coded and transmitted.
To reconstruct the block, the inverse transformation is applied, which, in the case of an orthonormal matrix, is provided by:
\begin{equation}
    \hat{\mathbf{x}} = Q\cdot\hat{\mathbf{y}}\mathbf{M}^{T}. 
\end{equation}

The objective is now to learn an optimal matrix $\mathbf{M}$ that, when applied to a set of residual blocks $\mathbf{x}$, generates transformed coefficients $\mathbf{y}$ which form a compact representation, minimizing both distortion and associated rate. With the rise of deep learning and end-to-end learned compression, the $\mathbf{M}$ matrix is simply learned through back-propagation and stochastic gradient descent. 

\begin{figure}
    \centering
    \includegraphics[width=0.9\linewidth]{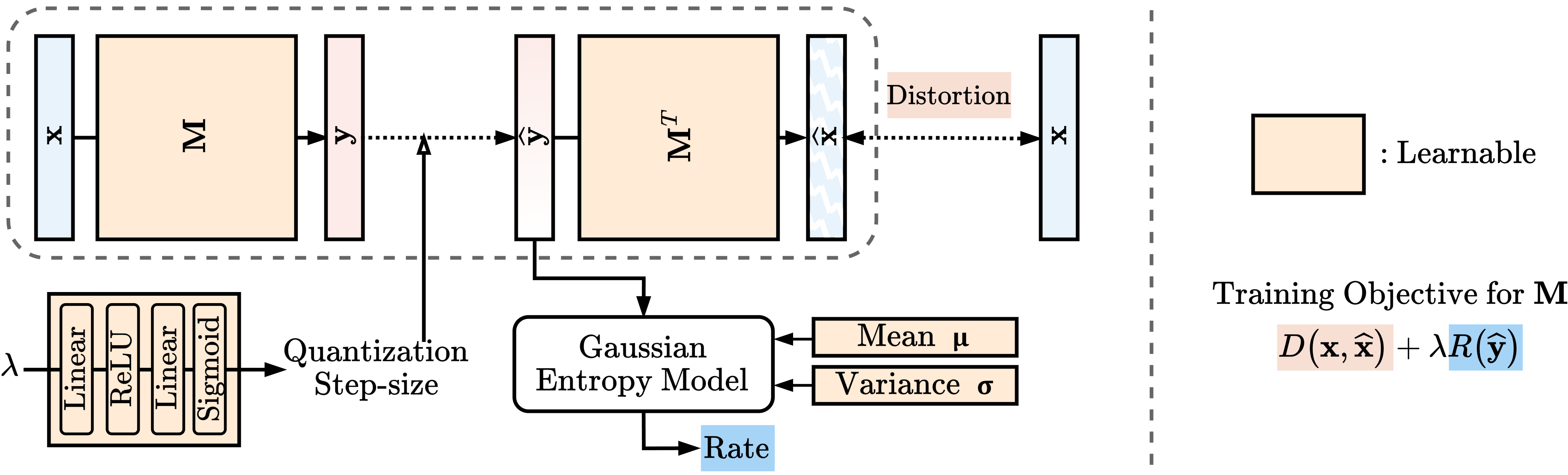}
    \caption{Learning framework of the proposed RDLT.}
    \label{fig:learning}
\end{figure}

\SubSection{Learning procedure}
The overall learning procedure is depicted in Figure~\ref{fig:learning}.
The transform matrix $\mathbf{M}$, which can be interpreted as a one-layer neural network with $n^4$ parameters, is trained to minimize the RD loss function:
\begin{equation}
    \mathcal{L} = D(\mathbf{x},\hat{\mathbf{x}})+\lambda R(\hat{\mathbf{y}}),
\end{equation}
where $D(\mathbf{x},\hat{\mathbf{x}})$ is the distortion term, calculated as the mean square error (MSE) between the original residual block $\mathbf{x}$ and the reconstructed one $\hat{\mathbf{x}}$, $R(\hat{\mathbf{y}})$ represents the rate term, which is the cost in bits per pixel required to encode the quantized coefficients $\hat{\mathbf{y}}$, and $\lambda \ge 0$ is the Lagrange multiplier determining the RD working point. To generate $\hat{\mathbf{x}}$, we utilize the transposed matrix $\mathbf{M}^T$ instead of the inverse $\mathbf{M}^{-1}$. This choice enforces orthonormality on the matrix $\mathbf{M}$, which is a desirable property. The rate term is computed as the negative log-likelihood of the coefficients, i.e. $R(\hat{\mathbf{y}}) = -\log_2 p(\hat{\mathbf{y}})$, where $p(\hat{\mathbf{y}})$ is the (estimated) distribution of the quantized coefficients $\hat{\mathbf{y}}$. We perform back-propagation with $\mathcal{L}$ and apply gradient descent to update the matrix $\mathbf{M}$ directly. This approach enables the model to adjust the basis vectors in $\mathbf{M}$, allowing it to adapt to the statistical properties of the residual blocks from a RD perspective.

Note that to leverage back-propagation, the entire process is required to be differentiable. Accordingly, we adopt the standard practice of using additive uniform noise in place of hard quantization during training, as commonly done in end-to-end learned compression~\cite{ballé2017endtoendoptimizedimagecompression}. We then approximate $\hat{\mathbf{y}}$ by adding uniform noise to the coefficients $\mathbf{y}$, thus:
\begin{equation}
    \tilde{\mathbf{y}} = \frac{\mathbf{y}}{Q} + \epsilon, \quad\epsilon\sim\mathcal{U}(-0.5,0.5),
\end{equation}
where $\mathcal{U}(-0.5,0.5)$ represents the uniform distribution from $-0.5$ to $0.5$.

Since the goal is to learn a single optimal matrix across various bit-rate points, which correspond to different quantization step-sizes, we adopt a variable-rate training procedure. During training, different $\lambda$ values are used along with their corresponding quantization step-sizes $Q$ to work at various RD trade-offs. To determine the optimal $Q$ for a given $\lambda$, a simple neural network composed of two linear layers is included. The network predicts the optimal quantization step-size when inputting a $\lambda$ value, and it is jointly trained with the matrix $\mathbf{M}$.

\begin{figure}[t]
    \centering
        \subfigure[RDLT (Ours)]{
        \centering
        \includegraphics[width=0.4\linewidth]{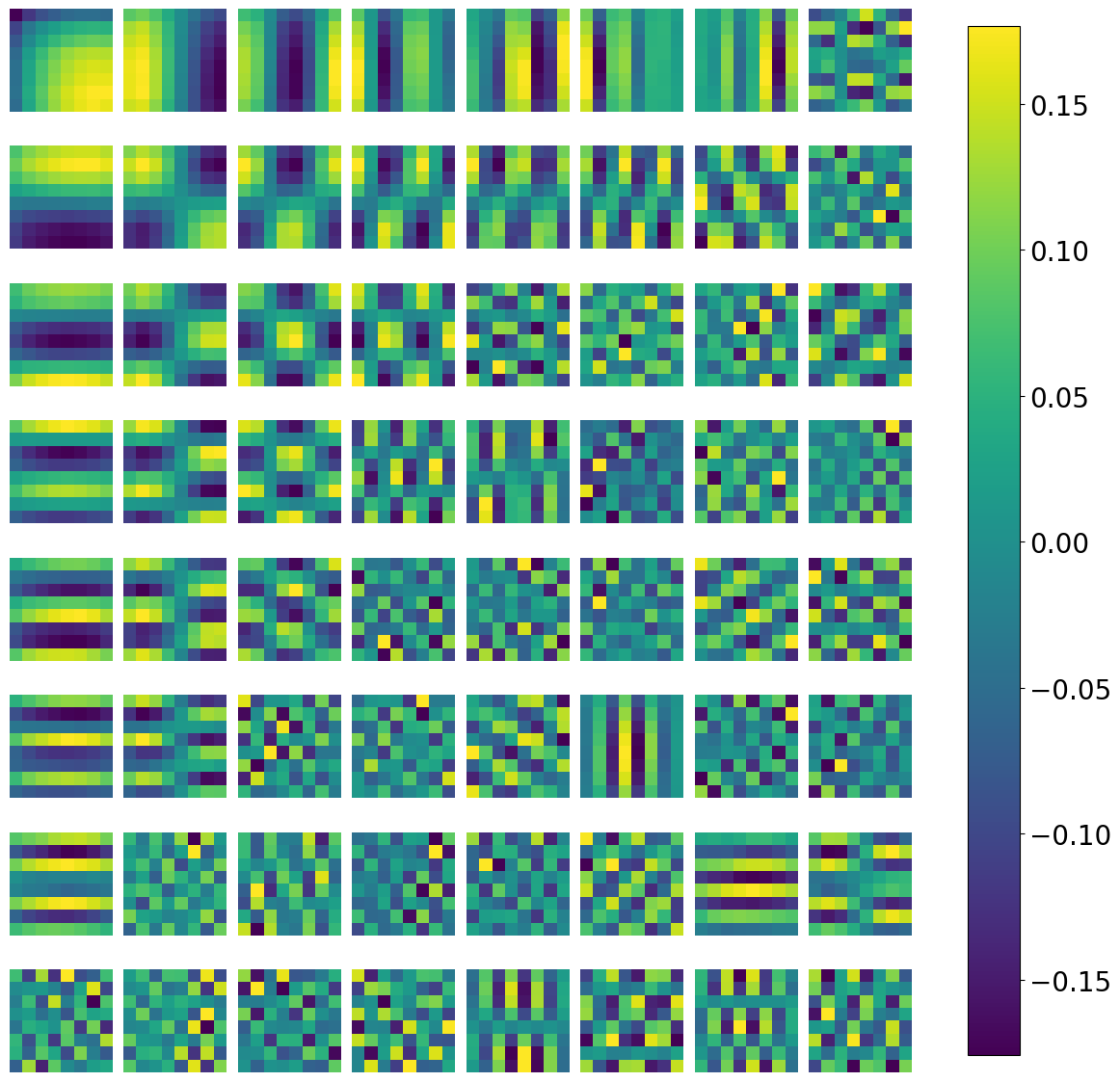}
        }
        \hspace{-0.2cm}
        \subfigure[DCT]{
        \centering
        \includegraphics[width=0.4\linewidth]{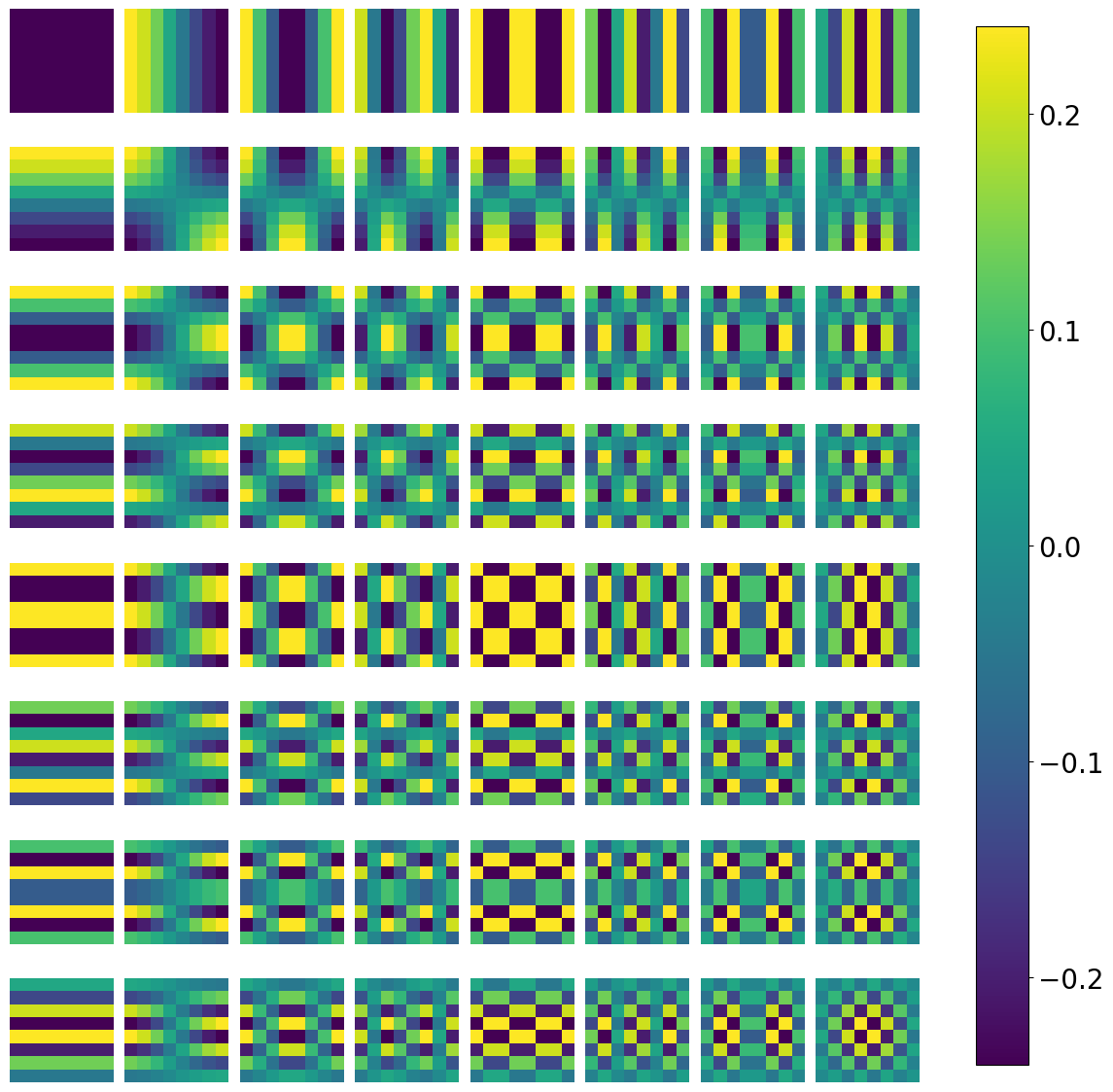}
        }
    \caption{Visualization of the basis functions of the RDLT and DCT with $n=8$.}
    \label{fig:vis_basis}
\end{figure}

\SubSection{Rate estimation}
Accurate rate estimation is essential for identifying the optimal transform from a RD perspective. Since the exact entropy of the transform coefficients is unavailable during training, we opt to use the Gaussian entropy model without hyperprior from~\cite{ballé2018variational} for estimation. Thus, we model each coefficient $\hat{y}_i$ as a Gaussian distribution convolved with the standard uniform distribution. In symbols, we have
\begin{equation}
    p(\hat{y}_i | \mu_i, \sigma_i) =  \mathcal{N}(\hat{y}_i | \mu_i, \sigma_i^2) \ast \mathcal{U}(-0.5,0.5),
\end{equation}
where $\mu_i$ and $\sigma_i$ denote the mean and variance of the Gaussian, which are learnable parameters trained during the optimization as well. Accordingly, we learn additional $2 n^2$ parameters, $n^2$ representing the means and $n^2$ representing the variances, which are then used to compute the entropy of the transform coefficients. The Gaussian entropy model offers a differentiable approximation of the true entropy, enabling end-to-end optimization of the rate term with the chosen loss function.  

Note that to adapt the entropy model for the variable-rate training procedure, where the distribution of the resulting coefficients $\tilde{\mathbf{y}}$ varies with different quantization step-sizes, the parameters of the Gaussian entropy model need to be adjusted as well. This is done by dividing the learned mean by $Q$ and the variance by $Q^2$ when calculating the entropy, ensuring a stable learning process.

\begin{figure}[t]
\begin{minipage}{0.55\textwidth}
    \centering
    \includegraphics[width=1\linewidth]{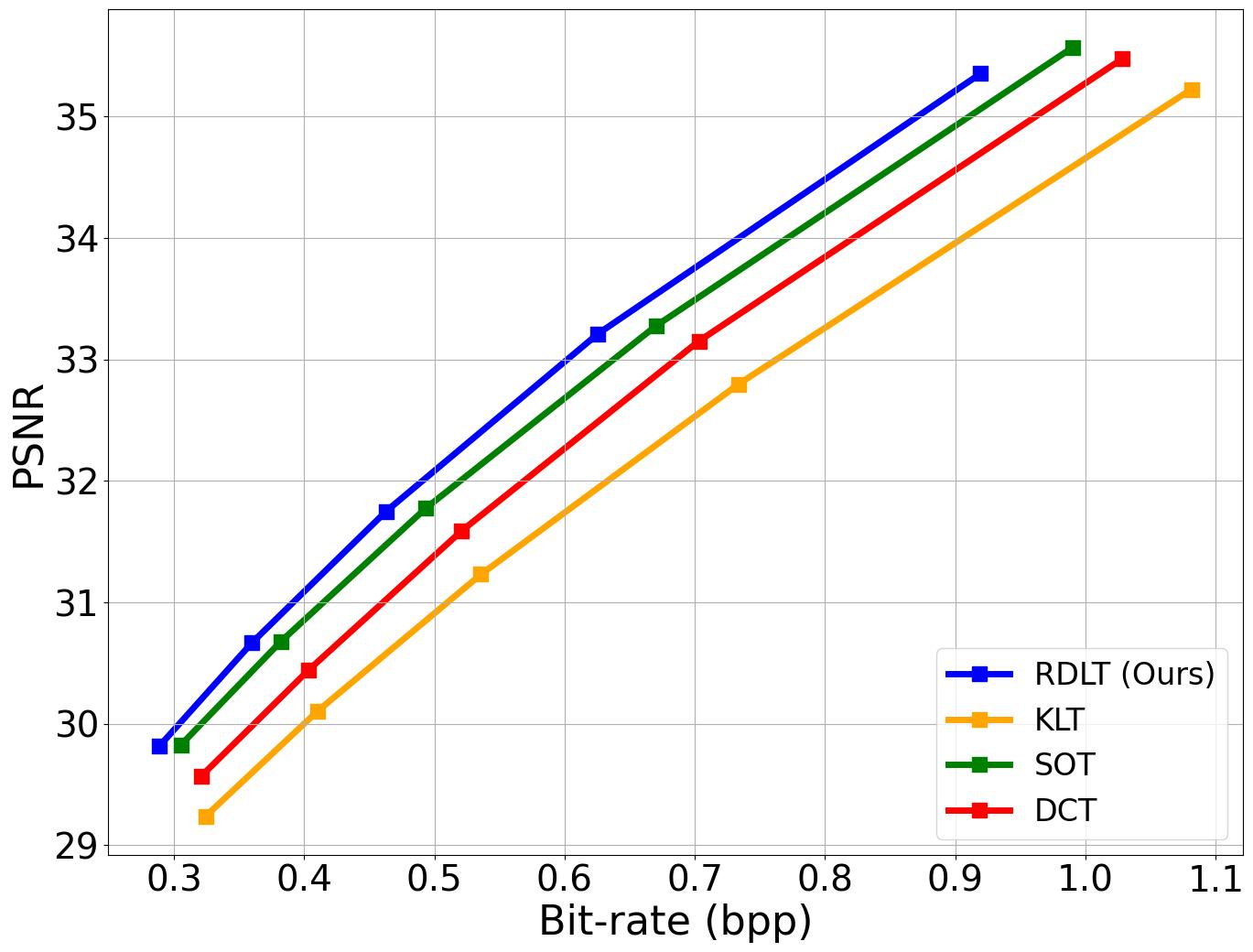}
    \caption{RD performance comparison.}
    \label{fig:rd_mix}
\end{minipage}
\hspace{2mm}
\begin{minipage}[b]{0.44\textwidth}
\small
\captionof{table}{BD-PSNR and BD-Rate comparison with DCT as the anchor.}
\begin{tabular}{@{}ccc@{}}
\toprule
            & BD-PSNR  & BD-Rate    \\ \midrule
DCT         & 0             & 0               \\
KLT         & -0.51         & 10.82           \\
SOT         & 0.41          & -7.82           \\
RDLT (Ours) & \textbf{0.68} & \textbf{-12.75} \\\bottomrule
\end{tabular}
\label{tab:BD_mix}
\end{minipage}
\hfill
\end{figure}

\Section{Experiments}
\label{sec:exp}

\SubSection{Experimental settings}
To validate the effectiveness of the RDLT for compression efficiency, a dataset containing residual blocks of several block sizes has been constructed. The source images are collected from a mix of image or video datasets for enriching the diversity, which includes a subset of ImageNet-1K~\cite{deng2009imagenet}, USC-SIPI~\cite{usc-sipi}, and HEVC test sequences. We utilize the intra prediction of HEVC for generating the residual blocks from these input images, resulting in around 800,000 non-overlapping blocks of three different sizes $n=[8,16,32]$ in total. The dataset is then split into 85\% for training and the remaining 15\% for evaluation. 

In training, we found out that initializing the transform matrix $\mathbf{M}$ with DCT basis vectors facilitates faster convergence. We learn a separate transform matrix $\mathbf{M}$ for each block size by using the subset of the training data that includes only the blocks of that specific size (refer to Figure~\ref{fig:vis_basis} for a visual comparison between the basis functions of our learned matrix $\mathbf{M}$ and the DCT, for $n = 8$). We begin by training the transform at a fixed high rate point, followed by variable-rate training with $\lambda$ values uniformly sampled from $[0.5, 0.01]$.

For performance evaluation, we compare RDLT with DCT, KLT, and SOT~\cite{sezer2015sparse}. Both KLT and SOT are derived using the same training dataset as RDLT for each block size. Coding efficiency is based on the PSNR between the original and reconstructed residual blocks, and the rate used to encode the coefficients. To compute the rate, arithmetic coding for the quantized coefficients has been used.
Quantization step-sizes have been set as $Q = [20, 30, 40, 50, 60]$ to work across different rate points. The same setup is applied to all the transforms to ensure a fair comparison.

\begin{figure}[t]
    \centering
        \subfigure[8x8]{
        \centering
        \includegraphics[width=0.3\linewidth]{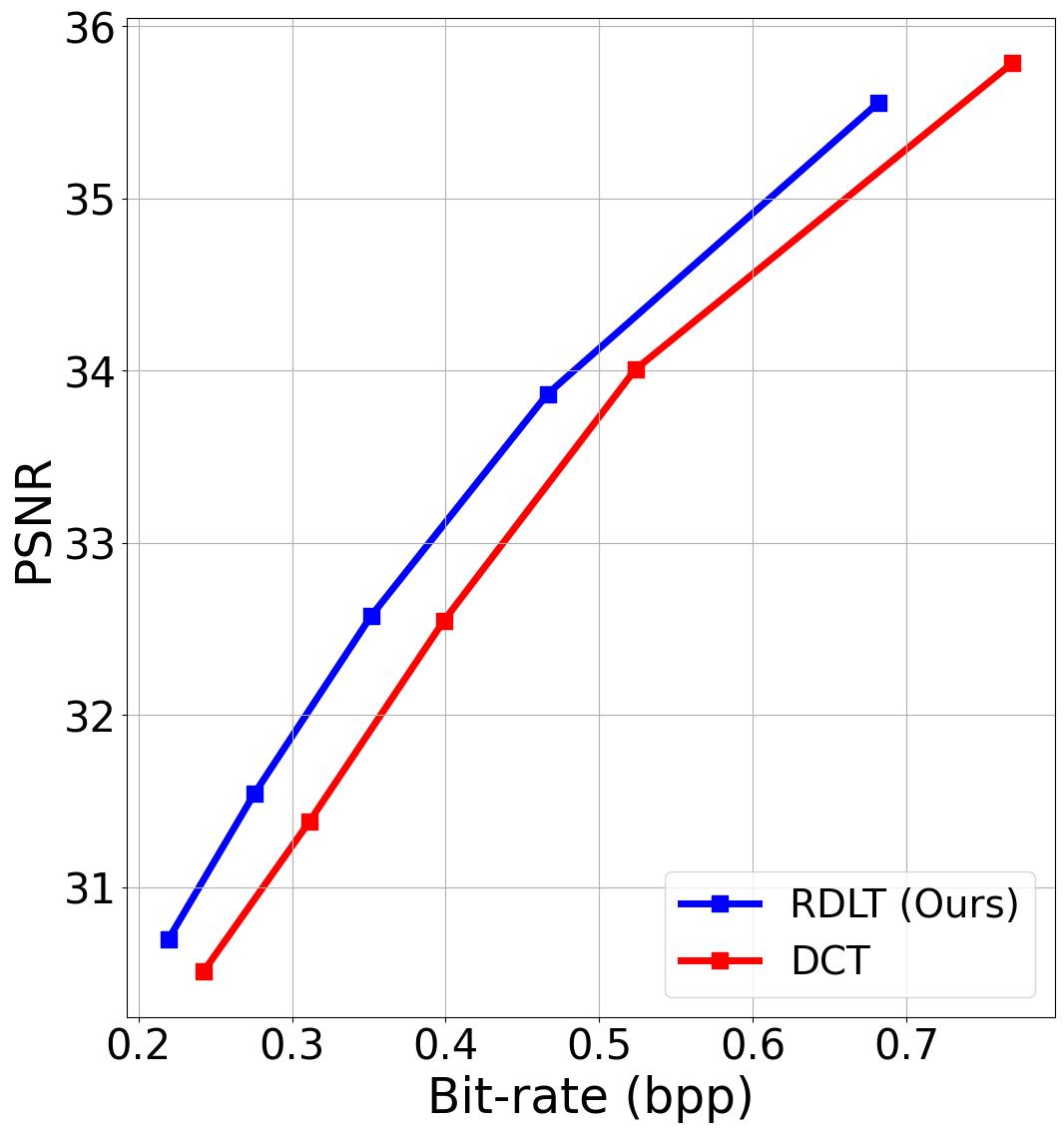}
        }
        \hspace{-0.2cm}
        \subfigure[16x16]{
        \centering
        \includegraphics[width=0.3\linewidth]{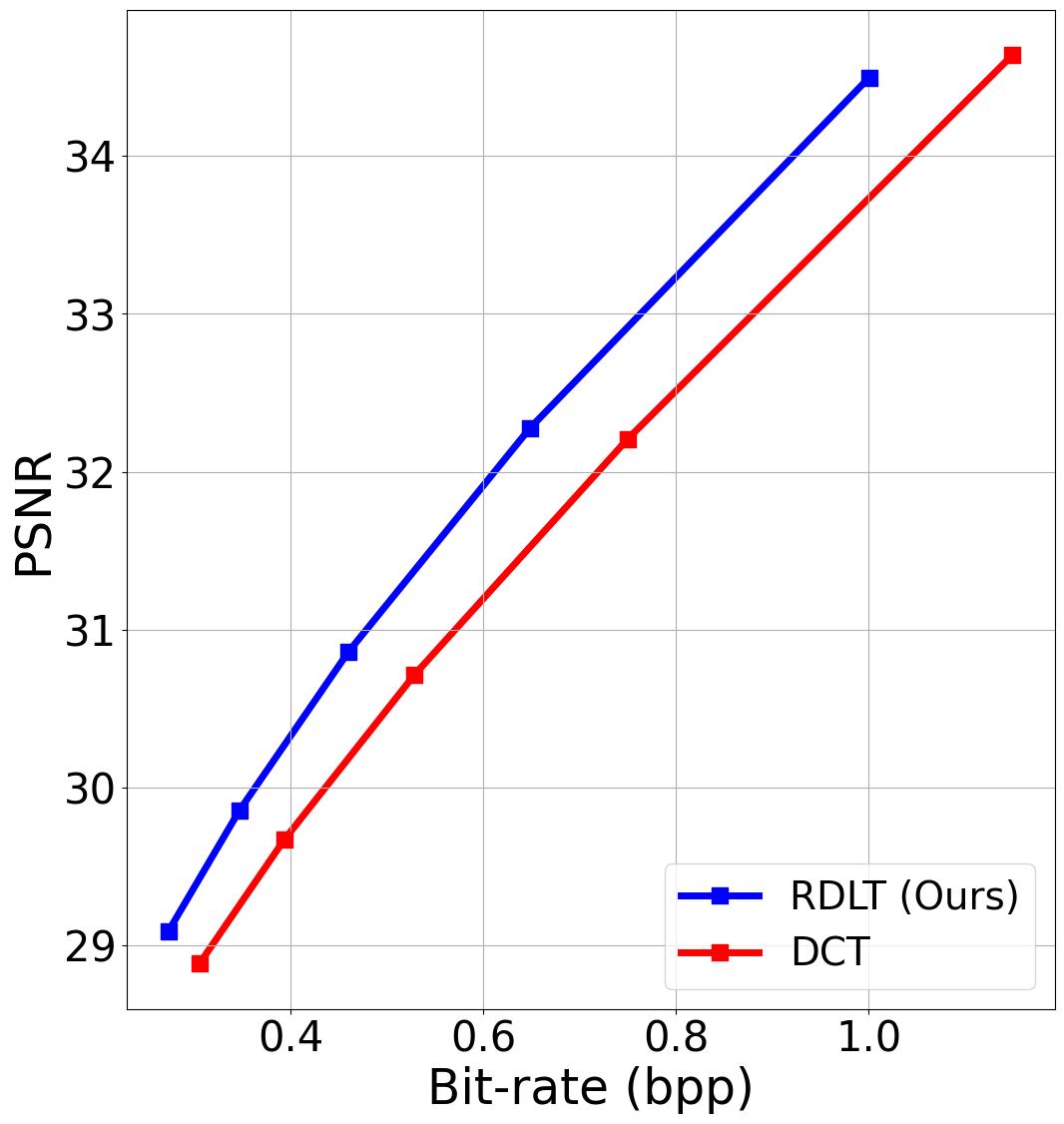}
        }
        \hspace{-0.2cm}
        \subfigure[32x32]{
        \centering
        \includegraphics[width=0.3\linewidth]{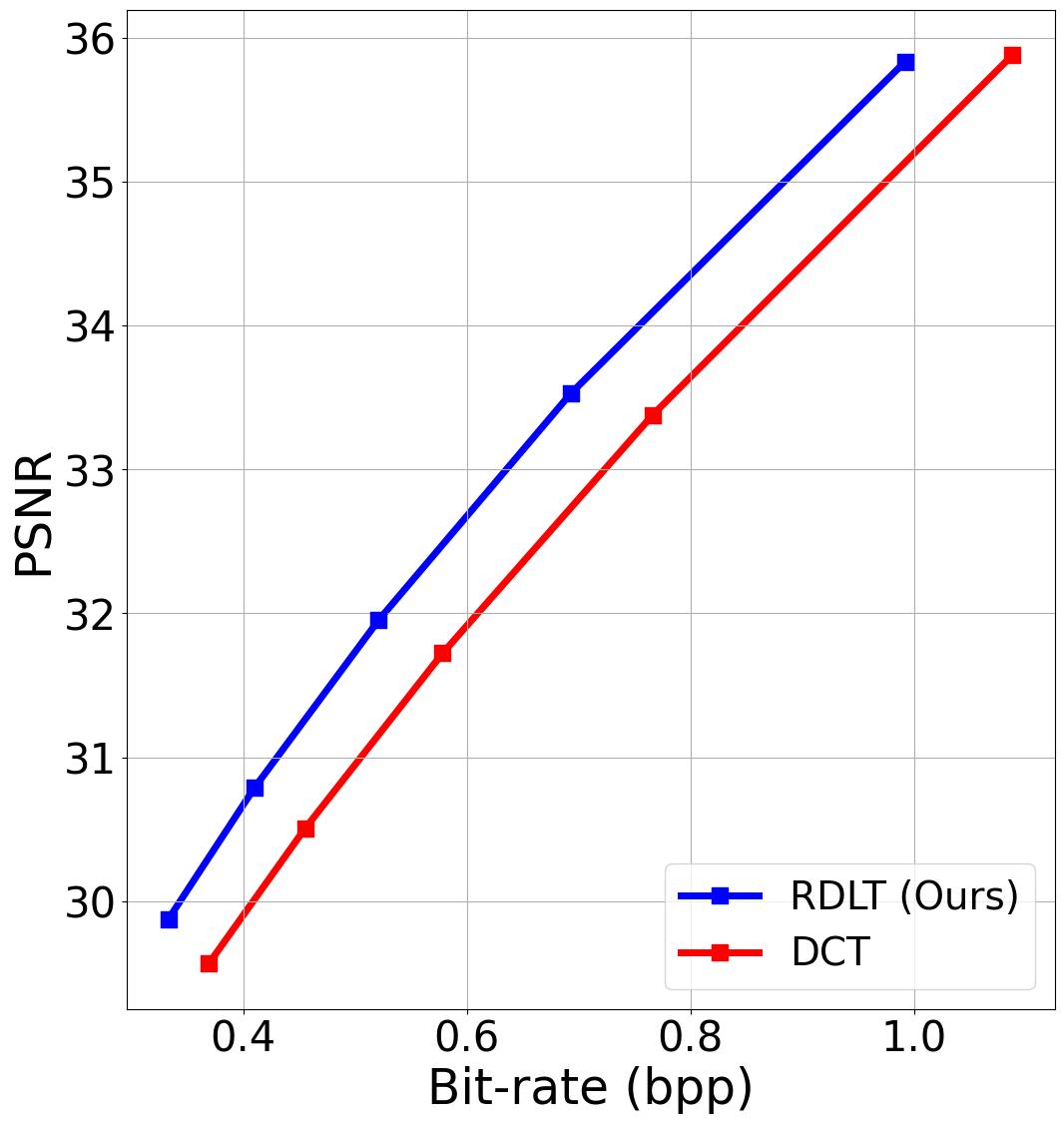}
        }
    \caption{Rate-distortion comparison for separate block sizes $n=[8,16,32]$.}
    \label{fig:rd_separate}
\end{figure}

\SubSection{Rate-distortion comparison}
Figure~\ref{fig:rd_mix} presents the RD comparison of RDLT and baseline transforms. The corresponding BD-PSNR and BD-Rate, with DCT as the anchor, are summarized in Table~\ref{tab:BD_mix}. The evaluation is conducted on the dataset that includes the combination of block sizes $n = [8, 16, 32]$. 

It can be observed that RDLT outperforms all the baseline transforms with a significant margin, particularly when compared to DCT. The proposed transform exhibits around $12\%$ BD-rate saving over the DCT, and $5\%$ over the SOT, which is optimized for nonlinear approximation but approximates the rate using the $l_0$ norm. This demonstrates the effectiveness of the end-to-end learned framework, particularly the use of the Gaussian entropy model for rate estimation. In contrast, the KLT, which is optimal for linear approximation, shows lower performance, indicating that its application in actual compression frameworks is far from optimal.

Figure~\ref{fig:rd_separate} provides a more in-depth view of RD performance across individual block sizes. The results indicate that RDLT outperforms DCT for each block-size with a consistent performance gain, supporting the robustness of the approach independently of the specific context in which it is applied.

\SubSection{Additional investigation}

\begin{figure}
    \centering
    \includegraphics[width=0.55\linewidth]{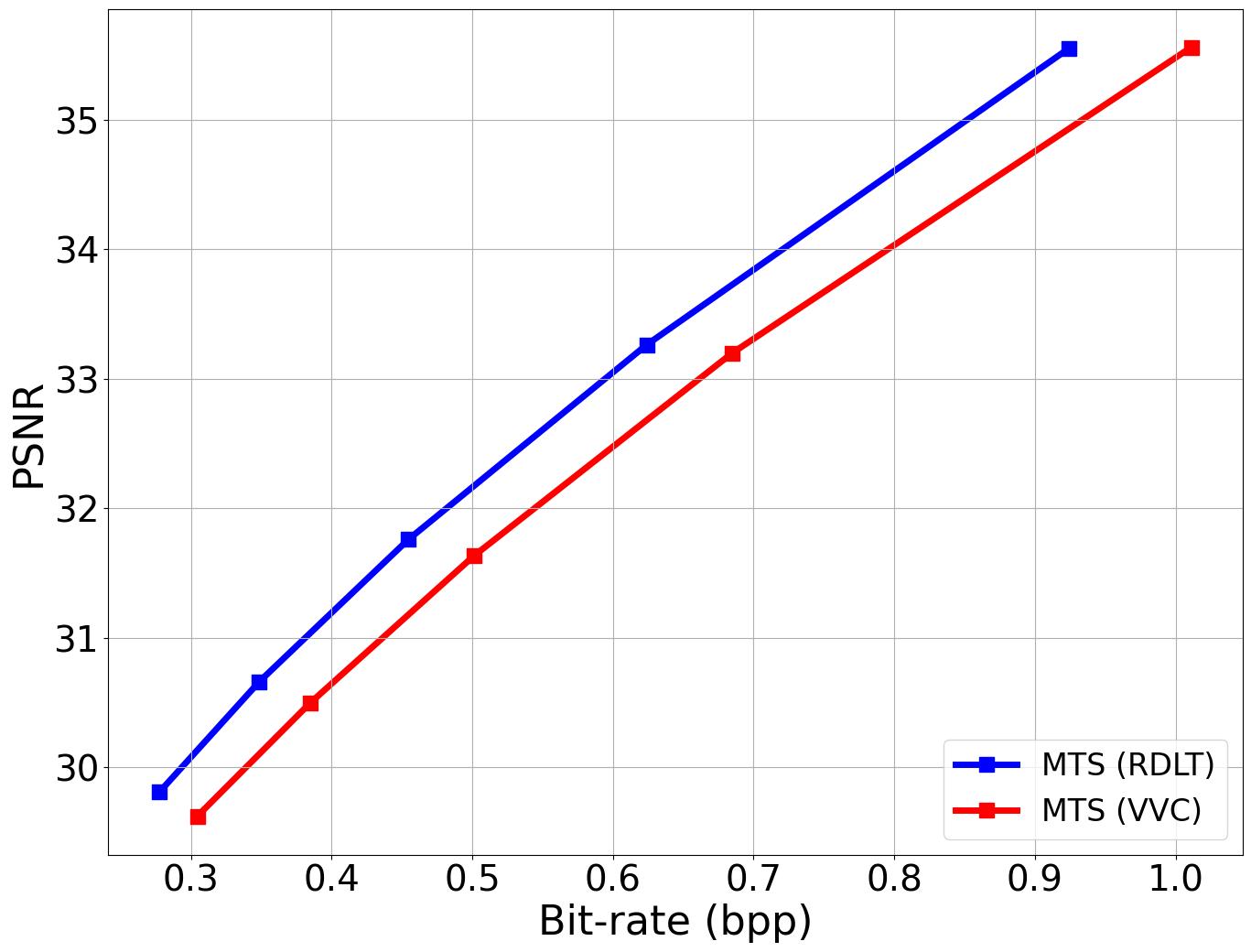}
    \caption{Rate-distortion comparison under multi transform selection in VVC.}
    \label{fig:rd_vvc}
\end{figure}

The Explicit Multiple Transform Selection (MTS) method in VVC enables the encoder to choose from various sinusoidal transforms based on the rate-distortion Optimization (RDO) paradigm. Specifically, VVC incorporates four additional primary transforms besides the DCT-II: the type-8 DCT (DCT-VIII), the type-7 DST (DST-VII), and their horizontal and vertical combinations. In this section, we simulate the integration of RDLT into the MTS framework of VVC. Figure~\ref{fig:rd_vvc} presents a performance comparison between the original MTS configuration in VVC and a modified version, in which the RDLT is integrated into the RDO process in place of the DCT. The results show that the system using the proposed transform surpasses the original MTS, confirming again its greater efficiency compared to DCT. This suggests that incorporating the RDLT into a real-world compression standard could significantly enhance overall performance in RD terms.

\Section{Conclusion}
\label{sec:conc}
In this paper, the Rate Distortion Learned Transform (RDLT) has been introduced. It represents a novel linear block transform designed to minimize the rate-distortion cost when approximating residual blocks resulting from standardized visual prediction used in current standards. The design of RDLT derives from advancements in end-to-end learned compression frameworks, leveraging modern learning techniques, but ensuring compatibility with existing image and video compression standards. The experimental results show that the RDLT significantly outperforms DCT or other data-driven transforms. When simulating its integration into a VVC-like image compression framework, RDLT brings also significant improvements. Full integration of the RDLT into image/video compression standard remains an avenue for future work.

\Section{References}
\bibliographystyle{IEEEbib}
\bibliography{refs}

\end{document}